\documentstyle[12pt]{article}
\def\be{\begin{equation}}
\def\ee{\end{equation}}
\def\re#1{({\ref{#1}})}
\def\ci#1{{\cite{#1}}}

\def\ndt{{\noindent}}
\def\IM{{\bf M}}

\def\Tr{{\rm Tr}}
\def\H{{\rm H}}

\def\pb{{\bar\partial}}

\def\b{{\beta}}
\def\d{{\delta}}

\def\ve{{\varepsilon}}
\def\m{{\mu}}

\def\l{{\lambda}}
\def\s{{\sigma}}

\def\o{{\omega}}

\def\CA{{\cal A}}

\def\CE{{\cal E}}
\def\CF{{\cal F}}

\def\CH{{\cal H}}

\def\CM{{\cal M}}
\def\CN{{\cal N}}
\def\CO{{\cal O}}

\def\CV{{\cal V}}

\def\IP{\bf P}

\def\IZ{\bf Z}

\begin{document}
\bigskip\bigskip
\centerline{\bf FRECKLED INSTANTONS IN TWO AND FOUR DIMENSIONS}
\bigskip
\centerline{A.~Losev $^{a}$, N.~Nekrasov$^{b}$,
S.~Shatashvili $^{c}$\footnote{On leave of
absence from Steklov Mathematical Institute, St.-Petersburg, Russia}} 
\medskip\centerline{$^{a,b}$ \it
Institute for Theoretical and
Experimental Physics, Moscow,  Russia}
\centerline{$^{b}$ \it Physics Dept., 
Princeton University, Princeton NJ 08544 USA}
\centerline{$^{c}$ \it Physics Dept., Yale University,
New Haven CT 06540 USA}
\bigskip
{\small Field theory with 
instantons can be partially
regularized by adding degrees of freedom at some scale. These
extra degrees of freedom lead to the appearence of the new
topological defects. These
defects which we call freckles have some characteristic size
depending on the scale at which the extra degrees of freedom
revive. 

The examples of two dimensional sigma model, 
four dimensional gauge theory are studied. The compactification
of the four dimensional supersymmetric gauge theory down
to two dimensions is also considered and the new phenomena are found.}

\vskip 1cm
\noindent {\tiny
Contribution to the Proceedings of Strings'99 Conference, Potsdam, 
July 1999}
\bigskip

\noindent\hbox{ITEP-TH-61/99}

\bigskip

\section{Introduction}

{}Two dimensional
sigma models and four dimensional gauge theories are similar
in many ways\ci{polyakovi}.
The interesting aspect of the both theories is the presence
of instantons, i.e. finite action solutions of the theory in
the Euclidean space-time.
Also, in both theories  the conformal invariance of the
instanton equations leads to the phenomenon of the
instanton shrinking. In this way the instanton moduli space
extends into the ultraviolet regions of the field space where
the degrees of freedom of more fundamental theory  may appear.

These degrees of freedom effectively compactify the instanton
moduli space, by adding the smooth field configurations which
look as singular point-like objects
from the point of view of low-energy  observer.
Their contribution to the correlation functions
is the phenomenon in the intersection theory known as
``the contribution of the connected component''
(sect. $9.1, 10.4$ in \ci{fulton}). Its subtraction
is an impact of the RG flow and is more or less contained in the
paradigm of integrating out of the  fast degrees of freedom
\ci{NSVZ, witph}.

In this short note we discuss briefly the concrete examples 
of this phenomena with implications to the studies
of supersymmetric gauge theories in four dimensions.

\section{Two dimensional instantons with freckles}

\ndt {\sl Non-linear sigma model.}
Consider the ${\CN}=2$ supersymmetric
sigma model (NLSM)
with the target space $V = {\CA}//G$ -- the K\"ahler
quotient of the K\"ahler  vector (affine) 
space $({\CA}, {\o})$ where $G$ acts preserving $\o$.
It has a moduli space ${\CM}$
of BPS field configurations - instantons,
represented by the holomorphic maps ${\Phi}$ of the
worldsheet $\Sigma$ into $V$. 
This space ${\CM}$ splits as a disjoint union of the spaces
${\CM}_{\b}$ labelled by ${\b} = \left[ {\Phi}({\Sigma}) \right] \in
{\H}_{2}(V,{\IZ})$. 
The complex (virtual)
 dimension of ${\CM}_{\b}$ is given by:
\be
{\rm dim}{\CM}_{\b} = \int_{\b} c_{1}(V)  + {\rm dim}V ( 1- g) 
\label{vdi}
\ee
with $g$ being the genus of $\Sigma$.

\ndt {\sl Gauged linear sigma model.}
One can construct a gauged linear sigma model (GLSM),
with chiral multiplets $\phi + \l + \ldots$ taking values in $A$, 
and vector multiplets $A + \psi + \ldots +\sigma$ taking values in Lie$G$.
The action of GLSM contains the potential term 
${\CV} = \langle \mu,  \mu \rangle$ where $\mu : {\CA} \to {\rm Lie}^{*}G$ 
is a moment map and $\langle , \rangle$ is a Killing form ($D$-term).  

At low energies the slowly varying field configurations 
are localized near the zeroes of $\mu$ and therefore describe
the maps to
$V ={\mu}^{-1}(0)/G$. 

However, one may be interested in BPS field configurations, 
and these obey the following equations \ci{witph}:
\be
F_{A} =  - e^2 \m, \quad {\pb}_{A} \phi = 0 
\label{bps}
\ee
with $e$ being the gauge coupling.

\ndt {\sl Topological classification.}
The moduli space ${\widetilde\CM}$ of BPS fields splits as a disjoint
union of the spaces ${\widetilde\CM}_{\d}$
corresponding to the topologically
distinct
gauge field configurations. The latter are labelled by
${\d} \in 
{\H}^{2}({\Sigma}, {\pi}_{1}(G)) \approx 
{\H}_{1}(G)$. 
In general there is no isomorphism between 
${\H}_{1}(G)$ and ${\H}_2 (V, {\IZ})$. As a consequence,
 the BPS
configurations in the GLSM and NLSM are labelled by different
groups. However, the canonical principal
$G$-bundle over $V$ gives an element (characteristic class)
of the group
$c \in 
{\H}^{2}(V, {\H}_{1}(G))$
and we get a map\footnote{It is the
ordinary topological rather then algebraic 
homology of the group $G$ that we use throughout this paper}: 
${\H}_2(V, {\IZ}) \to {\H}_1 (G)$:
\be
{\b} \mapsto {\d} = 
 \int_{\b} c \in {\H}_{1}(G)
\label{pair}
\ee

\ndt {\sl Holomorphic data.} The second equation in 
\re{bps}
 tells
us that $\phi$ is a holomorphic section of a vector
bundle ${\CE}$, whose holomorphic structure is specified
by ${\pb}_{A}$.
The space of such sections is denoted by ${\H}^{0}({\CE})$ - it is
a  vector (affine, finite-dimensional) space
for vector (affine, finite-dimensional) $\CA$. 
One has Riemann-Roch formula:
$$
{\rm dim}{\H}^{0}({\CE})  - {\rm dim}{\H}^1 ({\CE}) = d +
{\rm rk}{\CE}(1-g)
$$
where
 $d = {\rm deg}{\CE} = 
{1\over{2\pi i}}
\int_{\Sigma} {\Tr}F_{A}, \quad {\rm rk}({\CE}) = {\rm dim}{\CA}$.
If there is a bound
${\Tr} i \m \geq - r$ then 
only for $d \leq 
2\pi e^2 r $ that \re{bps} have solutions
(this is the sigma model analogue of the Donaldson
jumps in four dimensional gauge theory).

The space ${\CA}$ is acted upon the complexification $G_{c}$ of $G$. 
The potential ${\CV}$ viewed as a function on ${\CA}$
induces a gradient flow with respect to the K\"ahler metric
on ${\CA}$ along which the value of the potential decreases.
The set ${\CA}^{s}$ of points which flow to the zero set of ${\CV}$ is
distinguished for 
one has: $V = {\CA}^{s}/G_{c}$. 
The set of critical points of ${\CV}$ may have several connected components.
For a coadjoint orbit $\l$ of $G$ denote
by $C_{\l}$ the set of critical points $x$
of ${\CV}$ such that
${\mu}(x) \in {\l}$, and by $S_{\l}$ the points $y$ which flow
to $C_{\l}$ under the paths of the steepest
descend of $\CV$. The set of $\l$ for which $C_{\l}$ is not
empty is denoted by $\Lambda$.

The spaces
 ${\CH} = 
{\H}^{0}({\CE})$ and ${\CH}^{\prime} = {\H}^{1}({\CE})$
form a pair of vector bundles (actually their difference is
a better defined object as both may change dimension)
over the moduli space ${\IM}_{\Sigma}$ of holomorphic bundles
on $\Sigma$. Actually, not every holomorphic bundle is good.
More precisely, the equation $F_{A} = -e^2\mu$ picks
out only stable pairs $({\pb}_{A}, \phi)$ \ci{stable}. The quotient
of the space of stable pairs by the action of the complexified
gauge group is the moduli space of the BPS configurations.
It is simple to evaluate its dimension assuming that ${\CH}^{\prime} = 0$.
Since the complex dimension of ${\IM}_{\Sigma}$ equals
${\rm dim}G ( g- 1)$ and ${\rm dim}V = {\rm dim}{\CA} - {\rm dim}G$
then:
\be
{\rm dim}{\widetilde\CM}_{\d} =  \langle \d \rangle
 + {\rm dim}V ( 1- g)
\label{vdii}
\ee
If
${\d} = \int_{\b}c$ for ${\b} \in {\H}_{2}(V, {\IZ})$
then $\langle {\d} \rangle = \int_{\b} c_{1}(V)$. 

{}From now on put $g=0$. 
In this case ${\IM}_{\Sigma}$ is empty and the holomorphic bundles
have a 
stabilizer isomorphic to 
$G_{c}$. The space ${\CH}$ 
is acted upon by the
group $G_{c}$. The subset of ${\CH}$
which consists
of the points whose $G_{c}$ orbit intersects the space of
solutions to the equation $F_{A} = - e^2 \mu$ is denoted by
${\CH}^{s}$. The space of BPS field configurations is therefore
${\widetilde\CM} = {\CH}^{s}/G_{c}$.
The moduli space 
${\widetilde\CM}$ contains a subset ${\CM}$ of gauge
equivalence classes of sections whose value at each point
of $\Sigma$ belongs to ${\CA}^{s}$. The $\phi$'s from the
complement
${\widetilde\CM} \backslash {\CM}$ violate this condition
at some points $z_1, \ldots, z_k \in \Sigma$, where they
take values in $S_{{\l}_1}, \ldots, S_{{\l}_k}$. 

\ndt
{\sl Remark.} There is an interesting interplay between the 
$G$-equivariant
cohomology of ${\CA}$, cohomology of $V$, the 
set $\Lambda$, and the stratification of ${\widetilde\CM}$ \ci{kirwan}. 
$\widetilde\CM$ as (partial) compactification of $\CM$ was
studied in \ci{drinfeld, giv, plmor, witph, witgra}. 
It differs from the compactification 
by {\sl stable maps} 
\ci{km}. 

\ndt 
The points of $\widetilde\CM$ are called {\sl freckled instantons},
$z_1, \ldots, z_k$ are {\sl 
the freckles} of the types
${\l}_1, \ldots, {\l}_k$ respectively. In physics literature
they appeared as vortices \ci{ano, witph}.

\ndt {\sl The size of the freckles.}
As a smooth 
solution to \re{bps} the freckles of 
the type $\l$
have a size ${\ell}_{\l} \sim {1\over{e \sqrt{ \langle \l , \l
 \rangle}}}$ where we used the Killing form on Lie$^{*}G$.

\ndt {\sl Compactness of ${\widetilde \CM}$.} If $\Lambda$ is finite then
the number of extra topological defects one has added to the sigma model
instantons  is finite and their size is bounded from below.
One expects $\widetilde\CM$ to be compact in this case (of course
assuming the compactness of $V$). However, if $\Lambda$ is
infinite the space $\widetilde\CM$ is non-compact.

\ndt{\sl Correlation functions.} In NLSM
with the target space $V$ an interesting set of correlation functions
is obtained as follows: for $C \subset V$ and $z \in \Sigma$ define
the subsets\footnote{These subsets may have
self-intersections and cusps}
${\CM}_{C}^{(2)}, {\CM}_{C}^{(0)} (z) \subset {\CM}$
as follows: ${\CM}_{C}^{(2)}$ consists of such holomorphic maps
$\Phi: \Sigma \to V$ that ${\Phi} (\Sigma)$ intersects $C$;
${\CM}^{(0)}_{C}(z)$ consists of holomorphic maps $\Phi$
such that ${\Phi}(z) \in C$. 
Take the submanifolds $C_1, \ldots, C_k
 \subset V$, let $\omega_1, 
\ldots, \omega_k$ be the Poincare duals (delta functions supported at)
to $C_1, \ldots, C_k$ homology classes. Let $0 \leq p \leq k$.
Then the correlations functions of 
the observables ${\CO}^{(0)}_{\omega_1}(z_1), \ldots, 
{\CO}^{(0)}_{{\o}_p}(z_p),
\int
{\CO}^{(2)}_{{\o}_{p+1}}, \ldots \int
{\CO}^{(2)}_{{\o}_k}$ in the sigma model (see \ci{wittop} 
for their definition)
is equal to
the intersection number:
\be
\# {\CM}^{(0)}_{C_1}(z_1) \cap {\CM}^{(0)}_{C_p}(z_p)
\cap {\CM}^{(2)}_{C_{p+1}} \ldots \cap {\CM}^{(2)}_{C_{k}}
\label{inter}
\ee
In GLSM the corresponding correlation functions 
count
the intersection numbers of the closures ${\widetilde\CM}^{(0)}_{C}(z),
{\widetilde\CM}^{(2)}_{C}$ of the submanifolds
${\CM}^{(0)}_{C}(z), {\CM}^{(2)}_{C}$ in ${\widetilde\CM}$.
In general they differ considerably as the submanifolds
${\widetilde\CM}_{C}$ tend to intersect over the set of instantons
with freckles \ci{golfand}. 

So, are the freckles a nuisance or can they be helpful?
One answer to this question is suggested in \ci{golfand}
where it is shown that in principle one could
subtract the contribution of the freckles to get the
genuine answers from NLSM. 
Or, in certain correlation functions the freckles are harmless.
For example, if $c_1(V) >0$ then freckles do not
contribute to the 
quantum cohomology ring.

Finally, as we shall see in the next section, the presence
of freckles can be
{\sl necessary} in order to build up a higher-dimensional
theory.

\section{Four dimensional theory on a product of two Riemann surfaces}
\medskip
Take a genus
$h$ Riemann surface $C$,
 a 
compact semi-simple Lie
group $H$
 and consider the moduli space
of flat 
$H$-connections on $C$. To make things smooth
we assume that 
$H$ has a center $Z$ and consider
the flat connections in
 $H/Z$ bundle $E$ with non-trivial
$c_1 (E) \in {\H}_2 \left( C, {\pi}_1 (H/Z)\right)$ ('t Hooft
magnetic flux).
 This moduli space $V$ is an
infinite-dimensional symplectic quotient of the space $\CA$
of all $H$-gauge fields (we omit the details concerning $Z$ here
for brevity, they can be found in \ci{witdgt}) by the 
action of the infinite-dimensional group $G$ of $H$-gauge transformations.
The space $V$ has a canonical generator of ${\H}^{2}(V)$ - the
curvature of the determinant line bundle. We may take as the
K\"ahler form on $V$ the $k$-th multiple of this
generator. The number $k$ is called the level.

\ndt 
{\sl Four dimensional gauge theory emerges} if we were to
study the corresponding GLSM \ci{bln}. Indeed the matter fields
$\Phi$ are the gauge fields on $C$ which also
vary along the worldsheet $\Sigma$, the 
$G$
gauge fields are simply the gauge fields on $\Sigma$ which also 
vary along
$C$ so altogether we get an $H$ gauge field on $\Sigma \times C$.
The rest of the supermultiplets makes up a
four dimensional partially twisted ${\CN}=2$ vector multiplet.

\ndt {\sl Freckles.} Having learned the lessons of the previous
section we immediately raise the question of the interpretation of
the
 freckles. The BPS configurations in the GLSM turn out to 
be {\it precisely} four dimensional instantons on ${\Sigma \times C}$.
The two dimensional
gauge coupling $e^2$ is inversely
 proportional to
the
 area of $C$, $e^2 \sim
{1\over k
{\rm Area}_{C}}$, while the four dimensional coupling
is related to the level $k$ as $g^2 \sim {1\over k}$.
The labels ${\l}$ are interesting. They are nothing but the
dominant weights of $H$ and correspond to the higher
critical points of the Yang-Mills functional on $C$. 
The size of the type ${\l}$ freckle is therefore:
\be
{\ell}_{\l}^2 \sim {{k
{\rm Area}_{C}}\over{\langle {\l}, {\l} \rangle}}
\label{size}
\ee

\ndt {\sl Compactness of the instanton moduli space.} The fact that
one can
have the freckles of the 
arbitrarily small size 
is related to the four dimensional phenomenon of the instanton
shrinking. 
One way to cure this problem is to replace the instantons
by the torsion free sheaves, as in \ci{barth, gieseker, okonek, avatar,
nakajima}, or the non-commutative instantons \ci{noncomm}.
 ${\Sigma} \times C$
being K\"ahler allows that. There are new
phenomena which one encounters after this compactification is
performed. To be specific we study $U(2)$
 case. The classical instanton moduli space splits
as a $U(1)$ moduli space of flat connections on ${\Sigma} \times
C$ and an $SO(3)$ moduli space. Moreover the parity of
the $c_1$ of the $U(2)$ bundle is correlated with $w_2$ of the
$SO(3)$ bundle, while $U(1)$ bundle can have an abritrary 
$c_1$ allowed by the instanton equations. 

What happens when we hit a 
4d point-like instanton? If we are to treat this 4d freckle
in the holomorphic way
we should replace a (stable) holomorphic rank two bundle $\CE$
over the product of two curves $\Sigma \times C$ by the
torsion free sheaf $\CF$
which at some point ceases
to be locally free. The space of such sheaves 
(for 
fixed
 underlying holomorphic bundle 
${\CE}^{\circ}$ and fixed position 
$z \times p$ of the freckle) is isomorphic to 
a ${\IP}^1$ whose r\^ole
is
 the following. Take a holomorphic
bundle ${\CE}^{\circ}$ of the instanton charge less then that
of ${\CF}$ by one. 
Fix a complex line $l$ in the fiber ${\CE}^{\circ}_{z \times p}$
over a point $z \times p$. The space of such lines is a copy
of ${\IP}^1$ which we mentioned. Then the sheaf ${\CF}_l$ is a sheaf of
sections of ${\CE}^{\circ}$ whose value at the point
$z \times p$ belongs to $l$. 
One can show that 
the instanton charge of $\CF$ is a sum:
${\rm ch}_2
({\CF}) = 
{\rm ch}_2({\CE}^{\circ}) - 1,$
 (see \ci{golfand} for details on the
sheaves and freckles).

Now let us look at what happens from the point of view of the
two dimensional sigma model. We should restrict our sheaf
onto each fiber $w \times C$, $w \in \Sigma$. If the
restriction ${\CF}_w$ is stable then the
point $w$ is mapped to a point in the moduli space of
stable rank two 
bundles 
on $C$, i.e. for such a point $w$ we are  within the NLSM regime.
If for some point $w \neq z$ we get an unstable bundle ${\CF}_{w}$
then we get a two-dimensional
freckle -- the freckle of the GLSM.
The type $\l$ of freckle is presicely the type of unstable bundle
we got. 

Finally, if we hit the point $w = z$ then the sheaf we get
on $z \times C$ is a sheaf of sections of 
a vector bundle $E$
with smaller $c_1(E)$ then the first Chern class of the bundles
over nearby fibers!

So the message here is that the four dimensional
compactification adds in addition to the freckles which have
to do with the unstable bundles of the same topology as the
generic stable ones the bundles of different topology as well.
We hope to return to more detailed investigation of
this structure in future.

\ndt {\sl Quantum cohomology ring.} Suppose that $\Sigma \approx {\bf S}^2$
and take the limit where ${\rm Area}_{C} \ll {\rm Area}_{{\bf S}^2}$. 
We said above the quantum cohomology ring is not affected by
freckles, so we may compute the quantum cohomology of $V$ using 
four dimensional gauge theory. The manifold ${\Sigma} \times {\bf S}^2$
has $b_{2}^{+}=1$ which is the situation where
the Donaldson invariants get a contribution of the $u$-plane
integral \ci{moorewit} in addition to
 a Seiberg-Witten contribution. It can be computed using the
contact terms from \ci{issues} via the 
jumping techniques \ci{moorewit, mooremar}. The result is
interesting to compare with the results of the computations in another
chamber where $C$ is large and ${\bf S}^2$ is small in which case one
finds \ci{issues} a Landau-Ginzburg type topological sigma model
with the target space being a certain cover of the Seiberg-Witten
$u$-plane. In particular the topological ring is infinite-dimensional
in this case. To the contrary, the ring collapses once one
passes through all the jumping walls. We present the answer for the
$SO(3)$ case with $(w_{2}, [C]) \neq 0$. The classical cohomology ring
of $V$ is generated by the observables in the two dimensional Yang-Mills
 theory \ci{witdgt}: 
\be
a = \int_{C} {\CO}^{(2)}_{{\Tr}{\phi}^2}, 
\quad b = {\CO}^{((0)}_{{\Tr}{\phi}^2}
\label{gen}
\ee
$$
c = \sum_{i=1}^{h} \int_{A_{i}} {\CO}^{(1)}_{{\Tr}{\phi}^2} \int_{B^{i}}
{\CO}^{(1)}_{{\Tr}{\phi}^2}
$$
Then the answer stemming from the four dimensional twisted
${\CN}=2$ theory is:\footnote{the similar results
were obtained independently in \ci{lozmar}, although in a
less
concise form}
\be
\langle e^{{\ve}_1 a + {\ve}_2 b + {\ve}_3 c} \rangle \propto
\label{ans}
\ee
$$
\oint {{du dz}\over{(u^2-1)^{g} z^{g+1}}}
e^{2{\ve}_2 u + \left( {\ve}_1 u 
+ {\ve}_3 \left(u^2-1\right) \right)z}
{{{\s}_3({\ve}_1 +z)}\over{{\s}({\ve}_1){\s}_3 (z)}}
$$
where ${\s}_{3} (z) = 1 + {{u}\over{24}} z^2
+ \ldots$, ${\s}(z) = z + \ldots$ are the Weierstra{\ss} elliptic
functions associated to the SW
curve:
$$
y^2 = 4x^3 - {x \over 4} \left( {{u^2}\over 3} - {1\over 4} \right)
- {1\over 48} \left( {2u^3 \over 9} - {u \over 4} \right)
$$
The fact that the transcedental functions present in the similar
expression valid in the other chamber  cancel leading to the
finite dimensional quantum cohomology ring is a serious
check of the assertions above. It confirms the conjecture of \ci{bjsv}.

\ndt {\sl Refined Donaldson-Witten theory?.} 
In \ci{issues} it was argued that the
four dimensional gauge theory has much more interesting observables
then those studied in the Donaldson theory. In fact, to
each invariant polynomial $I({\phi})$ on the Lie algebra
of the gauge group one associates a descend sequence of observables, 
the highest term being the 4-observable which may be
used for deforming the action. On the other hand, using the fact
that the moduli space of flat connections is a Fano variety one can compute
its genus zero  Gromov-Witten invariants simply from
the WDVV equations using the reconstruction theory of \ci{km}.
But, as we have argued above, the four dimensional instantons
contain in addition to the two dimensional instantons also two
dimensional freckles of various types. So if we
were to compactify the four dimensional instanton moduli space
we would certainly get the space which looks very much different
from the moduli space of stable maps into $V$ above. Hence the
computation of Gromov-Witten invariants, although produces
the interesting function of the right number of
variables (if the curves $C$ of all genera are considered)
still misses the mark.

\ndt {\sl Higher dimensions.} In principle
we can construct the six-dimensional theory \ci{bln} at least at the level
of its BPS field configurations by taking a two dimensional Riemann
surface and studying its holomorphic maps into a moduli space
of instantons on a K\"ahler
fourfold $X$. Then, upon adding the freckles which are
connections obeying $F^{2,0}= 0$ and at the same time
reducible
stationary
points of the Yang-Mill functional we get a six dimensional BPS
gauge field. We believe these gauge fields are relevant in the
study of D1-D5 systems.

\centerline{\bf Acknowledgements}
\medskip\ndt
We thank H.~Braden, A.~Gorsky and A.~Rosly for discussions. 
The research of A.~L.~ is partly supported by  RFFI
under the grant 98-01-00328,
that of N.~N by R.H.Dicke Fellowship from Princeton University, partly by
RFFI under grant 98-01-00327; A.~L. and N.~N. are
partially supported by grant
96-15-96455 for scientific schools. Research of
S.~Sh.is supported
 by DOE grant
DE-FG02-92ER40704, by NSF CAREER award, by OJI award from
DOE and by Alfred
P.~Sloan foundation.

\end{document}